\documentclass[aps,nofootinbib,prd,eqsecnum,showpacs,showkeys,preprintnumbers]{revtex4-1}
\usepackage{graphicx}
\usepackage{graphicx}
\usepackage{amsmath}
\usepackage{amsfonts}
\usepackage{amssymb}
\usepackage{color}
\usepackage{bm}
\usepackage{float}
\usepackage{mathrsfs}
\usepackage{epstopdf}
\usepackage{url}
\usepackage{placeins}
\usepackage{footnote}
\usepackage{textcomp}
\usepackage[normalem]{ulem}
\usepackage[unicode=true, pdfusetitle,
 bookmarks=true,bookmarksnumbered=false,bookmarksopen=false,
 breaklinks=false,pdfborder={0 0 1},backref=false,colorlinks=false]{hyperref}
\usepackage{multirow}
\usepackage{pifont}
\usepackage{times}
\usepackage[english]{babel}

\usepackage{float}
\usepackage{enumerate}
\usepackage{lineno}
\usepackage{hyperref}

\usepackage{tabularx}
\makeatletter

\newcommand{\stkout}[1]{\ifmmode\text{\sout{\ensuremath{#1}}}\else\sout{#1}\fi}

\newcolumntype{L}[1]{>{\hsize=#1\hsize\raggedright\arraybackslash}X}%
\newcolumntype{R}[1]{>{\hsize=#1\hsize\raggedleft\arraybackslash}X}%
\newcolumntype{C}[1]{>{\hsize=#1\hsize\centering\arraybackslash}X}%

\newcommand*\patchAmsMathEnvironmentForLineno[1]{%
 \expandafter\let\csname old#1\expandafter\endcsname\csname #1\endcsname
 \expandafter\let\csname oldend#1\expandafter\endcsname\csname end#1\endcsname
 \renewenvironment{#1}%
   {\linenomath\csname old#1\endcsname}%
   {\csname oldend#1\endcsname\endlinenomath}}%
\newcommand*\patchBothAmsMathEnvironmentsForLineno[1]{%
 \patchAmsMathEnvironmentForLineno{#1}%
 \patchAmsMathEnvironmentForLineno{#1*}}%
\AtBeginDocument{%
\patchBothAmsMathEnvironmentsForLineno{align}%
\patchBothAmsMathEnvironmentsForLineno{flalign}%
\patchBothAmsMathEnvironmentsForLineno{alignat}%
\patchBothAmsMathEnvironmentsForLineno{gather}%
\patchBothAmsMathEnvironmentsForLineno{multline}%
}
\begin{document}

\title{Testing the bulk viscosity within the f(R,T) gravity in light of DESI-DR2 observations}
\author{Omayma Saddiki$^{1}$}
\email{omayma.saddiki@ump.ac.ma}
\author{Safae Dahmani$^{2, 3}$}
\email{dahmani.safae.1026@gmail.com}
\author{Taoufik Ouali$^{2, 3}$}
\email{t.ouali@ump.ac.ma}
\author{Abdelkarim Oukouiss$^{1}$}
\email{a.oukouiss@ump.ac.ma}

\date{\today }
\affiliation{$^{1}$Laboratory of Physical Sciences and Technological Innovation, 
Faculty of Applied Sciences of Nador, Mohammed I University, BP. 300, Selouane, 62702 Nador, Morocco\\$^{2}$Laboratory of Physics of Matter and Radiation, Mohammed I University, BP 717, Oujda, Morocco\\$^{3}$Astrophysical and Cosmological Center, Faculty of Sciences, Mohammed I University, BP 717, Oujda, Morocco}
\date{\today}
\begin{abstract}
We investigate the cosmological implications of the bulk viscosity within the framework of $f(R,T)$ gravity by considering the linear model $f(R,T)=R+2\lambda T$, where $\lambda$ characterizes the matter--geometry coupling. Assuming a homogeneous and isotropic Friedmann--Lema\^{i}tre--Robertson--Walker Universe, we derive analytical expressions for the Hubble parameter in two different scenarios: model A, corresponding to a pressureless viscous fluid ($\omega=0$), and model B, where the equation of state (EoS) parameter is treated as a free parameter. The cosmological parameters are constrained through a Markov Chain Monte Carlo analysis using the latest DESI-DR2 measurements, Pantheon$^+$+SH0ES compilation, and direct $H(z)$ observations. The statistical comparison of the proposed models is evaluated using the minimum chi-square $(\chi^2)$ together with the corrected Akaike Information Criterion (AICc), and the Bayesian Information Criterion (BIC). The results show that both viscous $f(R,T)$ models provide an excellent description of the current expansion history and remain compatible with the combined observational data. Model A is slightly favored over the standard cosmological model ($\Lambda$CDM) according to AICc, reflecting the improved fit achieved with only one additional free parameter. In contrast, the BIC continues to favor the simpler $\Lambda$CDM model owing to its stronger penalty on model complexity. Although model B yields the lowest minimum $\chi^2$, the improvement is too small to justify the introduction of an additional free parameter, resulting in a statistical performance comparable to $\Lambda$CDM according to the AICc but less favorable according to the BIC.
 Furthermore, the cosmographic analysis confirms that both models successfully reproduce the transition from an early decelerated phase to the present accelerated expansion while exhibiting moderate deviations from the standard $\Lambda$CDM predictions in the higher-order cosmographic parameters. These findings indicate that matter--geometry coupling together with bulk viscosity provides a viable extension of the standard cosmological framework and deserves further investigation with forthcoming high-precision cosmological observations.
\end{abstract}

\maketitle

\section{Introduction}
\label{sec1}
Modern cosmology has undergone remarkable progress over the last few decades owing to the continuous improvement of observational techniques and the increasing precision of astronomical surveys. One of the most significant discoveries in this field is the late-time accelerated expansion of the Universe \cite{Riess1998}, first revealed through observations of distant Type Ia supernovae (SNIa) \cite{Riess1998,Filippenko1998,Perlmutter1999} and subsequently confirmed by independent cosmological probes, including measurements of the Cosmic Microwave Background (CMB) \cite{Ade2016}, Baryon Acoustic Oscillations (BAO) \cite{eBOSS2021}, galaxy clustering \cite{Zehavi2011}, weak gravitational lensing \cite{Bartelmann2001}, and large-scale structure observations \cite{Eisenstein2005,Blake2012,Parkinson2012,Kazin2014,Alam2015,Beutler2016}. The remarkable agreement among these independent datasets has firmly established cosmic acceleration as one of the fundamental observational pillars of contemporary cosmology.

Within the framework of General Relativity (GR), the simplest explanation for the accelerated expansion is provided by the standard $\Lambda$CDM model \cite{Planck2018}, in which the Universe is composed primarily of cold dark matter (CDM) and a cosmological constant, $\Lambda$, that acts as dark energy with a constant equation of state (EoS) parameter equal to $-1$ \cite{Peebles2003}. Owing to its mathematical simplicity and excellent agreement with a wide range of cosmological observations, the $\Lambda$CDM scenario has become the standard paradigm for describing the large-scale evolution of the Universe. It successfully reproduces the observed expansion history, the anisotropy spectrum of the CMB, the formation of large-scale structures, and many other cosmological observables with only a small number of free parameters.

Despite these remarkable achievements, the physical interpretation of the cosmological constant remains one of the most challenging problems in theoretical physics. In particular, the enormous discrepancy between the observed value of the vacuum energy density \cite{Padmanabhan2003} and the prediction of quantum field theory gives rise to the well-known cosmological constant problem \cite{Weinberg1989}. Furthermore, the coincidence problem \cite{Steinhardt1999,Zlatev1999}, which questions why the energy densities of matter and dark energy are of the same order of magnitude only during the present cosmological epoch, remains unresolved. In addition to these theoretical issues, recent high-precision observations have revealed persistent tensions between independent determinations of several cosmological parameters, most notably the Hubble tension \cite{Riess2022,DiValentino2021,Freedman2023b,Adil2024b,Riess2024b,Dahmanisafae}, and $S_8$ tension \cite{Abdalla2022}. Although the origin of these discrepancies is still under investigation, they have stimulated considerable interest in exploring extensions of the standard cosmological model \cite{Dahmani2024,Mhamdi2024,Errahmani2026}.

One of the most promising directions consists of modifying the gravitational sector itself rather than introducing an unknown dark energy component. In modified gravity theories, the observed acceleration is interpreted as a consequence of deviations from Einstein's GR on cosmological scales. During the past two decades, numerous modified gravity theories have been proposed, including $f(R)$ gravity \cite{Starobinsky2007,Sotiriou2010,Sotiriou2009,DeFelice2010}, where $R$ denotes the Ricci scalar, $f(\mathcal{T})$ gravity \cite{Capozziello2011,Bamba2013,Koussourmd}, with ${\mathcal{T}}$ representing the torsion scalar, and $f(Q)$ gravity \cite{Mhamdi2024,Lazkoz2019,Mandal2020,Enkhili2024,Koussour2023}, in which $Q$ is the non-metricity scalar. Other important extensions include Gauss--Bonnet gravity \cite{Nojiri2005,Cognola2006,DeFelice2010GB}, scalar--tensor theories \cite{Brans1961,Faraoni2004,Clifton2012}, and several hybrid extensions \cite{Capozziello2015}. These models provide different geometrical mechanisms capable of reproducing the late-time accelerated expansion while simultaneously offering new perspectives on the nature of gravity beyond Einstein's theory.

Among these alternatives, $f(R,T)$ gravity has attracted particular attention because of its explicit dependence on both the Ricci scalar, $R$, and the trace of the energy--momentum tensor, $T$ \cite{Harko2011,Errahmani2026,Bouali2023,Sharif2012,Myrzakulov2023}. This additional dependence naturally introduces a non-minimal coupling between matter and geometry, leading to modified gravitational field equations and richer cosmological dynamics. The flexibility of this framework has motivated numerous investigations devoted to inflation, dark energy phenomenology, compact stars, anisotropic cosmologies, wormhole solutions, and other astrophysical applications \cite{Bouali2019,Dahmani2026,Alvarenga2013,Moraes2016666,Shabani2014,Deb2018}. Consequently, $f(R,T)$ gravity has become an active research area in the quest for viable alternatives to the standard cosmological model.

The cosmological implications of modified gravity become even more interesting when dissipative processes are incorporated into the cosmic fluid. In realistic physical systems, the assumption of a perfect fluid is generally an idealization, whereas irreversible thermodynamic processes naturally generate dissipative effects. Among the different types of viscosity, bulk viscosity plays a particularly important role in cosmology because it preserves the homogeneity and isotropy of the Friedmann--Lemaître--Robertson--Walker (FLRW) spacetime while introducing an effective negative pressure  \cite{Murphy1973,Maartens1995,Brevik2017}. Such a contribution can significantly modify the cosmic expansion history and has therefore been proposed as a possible mechanism for explaining the accelerated evolution of the Universe without invoking exotic dark-energy components.

The cosmological role of bulk viscosity has been investigated within both GR and several modified theories of gravity. Early studies \cite{Murphy1973,Maartens1995,Brevik2017} demonstrated that viscous effects may influence the evolution of the early Universe, alleviate cosmological singularities, and modify the transition between different expansion phases. More recently, bulk-viscous cosmological models have been employed to describe the late-time acceleration \cite{Cataldo2005,Velten2013,Brevik2020}, showing that dissipative processes may successfully reproduce several observational features of the expanding Universe. These investigations have highlighted that the viscosity coefficient represents an additional physical degree of freedom capable of affecting the background dynamics and the effective EoS of the cosmic fluid.

Within the framework of $f(R,T)$ gravity, viscous cosmology has received increasing attention in recent years \cite{Koussour2024}. Several authors have derived exact cosmological solutions by considering different functional forms of $f(R,T)$, various parameterizations of the bulk viscosity coefficient, and different assumptions concerning the cosmic matter content. These studies have shown that the coupling between matter and geometry, combined with dissipative effects, can generate a rich cosmological phenomenology, including accelerated expansion, modified deceleration histories, and viable dark energy behavior. Furthermore, observational analyses have demonstrated that many of these models remain compatible with current cosmological data \cite{Errahmani2026,Bouali2023,Koussour2024}, encouraging further investigations of viscous scenarios within modified gravity.

Despite these developments, several aspects remain insufficiently explored. In particular, many previous studies were based on earlier observational compilations or focused on specific cosmological fluids without performing a systematic comparison between different EoS. Moreover, the rapid improvement in observational precision, especially following the release of the Dark Energy Spectroscopic Instrument (DESI-DR2) \cite{DESI2025II,DESI2025I} together with the latest Pantheon+ Type Ia supernova compilation \cite{PantheonPlus2022,Scolnic2022} and the $H(z)$ measurements \cite{Jimenez2002,Moresco2012,Moresco2016}, provides a timely opportunity to re-examine viscous cosmological models under significantly tighter observational constraints. The inclusion of these recent datasets allows a more reliable determination of the free cosmological parameters and offers a stringent test of the viability of modified gravity models beyond the standard $\Lambda$CDM scenario \cite{Planck2018}.

Motivated by these considerations, the present work investigates cosmological models in the framework of $f(R,T)$ gravity \cite{Harko2011,Errahmani2026,Bouali2023,Sharif2012} by adopting the linear functional form $f(R,T)=R+2\lambda T$ \cite{Harko2011,Moraes2016,Myrzakulov2012}, where the parameter $\lambda$ quantifies the strength of the matter--geometry coupling. In addition to this coupling, the cosmic fluid is assumed to possess bulk viscosity \cite{Maartens1995,Murphy1973,Weinberg1971}, leading to an effective pressure that modifies the standard cosmological dynamics \cite{Brevik2005}. Two different cosmological scenarios are considered, model A where the cosmic fluid is described by pressureless matter ($\omega=0$), and model B where EoS parameter $\omega$ is treated as an additional free parameter and is constrained simultaneously with the remaining cosmological parameters. This approach makes it possible to investigate how relaxing the dust assumption influences both the cosmological evolution and the observational constraints.

The viability of the proposed models is examined by confronting their theoretical predictions with recent cosmological observations. The parameter estimation is performed through a Markov Chain Monte Carlo (MCMC) analysis \cite{Padilla2021} using the combined Pantheon+ sample \cite{PantheonPlus2022,Scolnic2022}, the latest DESI-DR2 measurements \cite{DESI2025II,DESI2025I}, and the $H(z)$ measurements \cite{Jimenez2002,Moresco2012,Moresco2016}. These complementary datasets probe different aspects of the expansion history of the Universe and provide stringent constraints on the free parameters of the models. Furthermore, the statistical performances of the proposed scenarios are evaluated using the minimum chi-square together with the Akaike Information Criterion (AIC) \cite{Akaike1974}, the corrected Akaike Information Criterion (AICc) \cite{Burnham2002}, and the Bayesian Information Criterion (BIC) \cite{Schwarz1978}, allowing a direct comparison with the standard $\Lambda$CDM cosmology.

The present analysis aims to address two closely related questions. First, we investigate whether the introduction of the bulk viscosity within the $f(R,T)$ framework remains compatible with the most recent cosmological observations. Second, we examine the impact of treating the EoS parameter as either fixed or free on the inferred cosmological parameters and on the statistical performance of the models. Rather than proposing a replacement for the standard cosmological model, our objective is to assess the observational viability of these generalized viscous scenarios and to quantify the role played by the matter--geometry coupling and the viscous effects in the late-time evolution of the Universe.

The remainder of this paper is organized as follows. In Sec.~\ref{sec2}, we briefly review the basic formalism of $f(R,T)$ gravity. Sec.~\ref{sec3} is devoted to the derivation of the cosmological field equations and the analytical solutions corresponding to models A and B. The observational datasets together with the statistical methodology are presented in Sec.~\ref{sec4}. The observational constraints and their cosmological implications are discussed in Sec.~\ref{sec5}. Finally, the main conclusions are summarized in Sec.~\ref{sec6}.

\section{Brief Review of \texorpdfstring{$f(R,T)$}{f(R,T)} Gravity}
\label{sec2}
The $f(R,T)$ theory represents an extension of GR in which the gravitational Lagrangian is allowed to depend not only on the Ricci scalar, $R$, but also on the trace of the energy--momentum tensor, $T$. Such a dependence introduces an explicit coupling between matter and geometry, leading to modified gravitational dynamics beyond Einstein's theory\cite{Harko2011}. Throughout this work, we adopt the system of units $8\pi G=c=1$, so that the gravitational action is written as
\begin{equation}
S = \frac{1}{2} \int d^4x \sqrt{-g} \left[f(R,T) + 2\mathcal{L}_m \right],
\label{2.1}
\end{equation}
where $g$ denotes the determinant of the metric tensor $g_{\mu\nu}$ and $\mathcal{L}_m$ is the matter Lagrangian density. The corresponding energy--momentum tensor $T_{\mu\nu}$ is defined through the variation of the matter action with respect to the metric according to
\begin{equation}
T_{\mu\nu} = -\frac{2}{\sqrt{-g}} \frac{\delta (\sqrt{-g}\mathcal{L}_m)}{\delta g^{\mu\nu}}.
\label{E2}
\end{equation}
Assuming that the matter Lagrangian depends only on the metric components and not on their derivatives, the energy--momentum tensor can be expressed in the form
\begin{equation}
T_{\mu\nu} = g_{\mu\nu}\mathcal{L}_m - 2 \frac{\partial \mathcal{L}_m}{\partial g^{\mu\nu}}.
\label{E2.3}
\end{equation}
Varying the action Eq.~(\ref{2.1}) with respect to the metric tensor yields the field equations of $f(R,T)$ gravity,
\begin{equation}
f_R(R,T)R_{\mu\nu} - \frac{1}{2}f(R,T)g_{\mu\nu}
+ (g_{\mu\nu}\Box - \nabla_\mu \nabla_\nu)f_R(R,T)
= T_{\mu\nu} - f_T(R,T)T_{\mu\nu} - f_T(R,T)\Theta_{\mu\nu},
\label{E4}
\end{equation}
where $f_R=\partial f/\partial R$ and $f_T=\partial f/\partial T$. Moreover, $\nabla_\mu$ denotes the covariant derivative associated with the metric, while $\Box=\nabla^\mu\nabla_\mu$ is the d'Alembert operator. The tensor $\Theta_{\mu\nu}$ is introduced through
\begin{equation}
\Theta_{\mu\nu} \equiv g^{\alpha\beta} \frac{\delta T_{\alpha\beta}}{\delta g^{\mu\nu}}.
\label{E2.5}
\end{equation}
By substituting the expression of the energy--momentum tensor, Eq.~(\ref{E2.3}), into the definition of $\Theta_{\mu\nu}$, Eq.~(\ref{E2.5}), and varying with respect to the metric tensor, we obtain
\begin{equation}
\Theta_{\mu\nu} = -2T_{\mu\nu} + g_{\mu\nu}\mathcal{L}_m
- 2g^{\alpha\beta} \frac{\partial^2 \mathcal{L}_m}{\partial g^{\mu\nu}\partial g^{\alpha\beta}}.
\label{E2.6}
\end{equation}
Several functional forms of $f(R,T)$ have been proposed in the literature in order to investigate different cosmological and astrophysical scenarios. In the present work, we focus on one of the simplest and most widely used choices (as one considered in \cite{Harko2011}),
\begin{equation}
f(R,T) = R + 2f(T).
\label{E7}
\end{equation}
This particular form preserves the Einstein--Hilbert term while introducing an additional contribution depending only on the trace of the energy--momentum tensor. Consequently, the modified gravitational field equations reduce to
\begin{equation}
R_{\mu\nu} - \frac{1}{2}Rg_{\mu\nu}
= T_{\mu\nu} - 2(T_{\mu\nu} + \Theta_{\mu\nu})f'(T) + f(T)g_{\mu\nu}.
\label{E8}
\end{equation}
To account for dissipative processes during the cosmic evolution, we assume that the matter sector is described by a bulk viscous fluid. Bulk viscosity provides an effective mechanism capable of modifying the thermodynamic pressure and has been extensively employed in cosmological models to describe irreversible processes in an expanding homogeneous and isotropic Universe. The effective pressure of the cosmic fluid is given by \cite{Brevik2005}
\begin{equation}
\bar{p} = p - \xi \theta = p - 3\xi H,
\label{E9}
\end{equation}
here, $p$ denotes the equilibrium pressure, $\xi > 0$ is the bulk viscosity coefficient, and $H$ is the Hubble expansion rate. The corresponding energy--momentum tensor of the viscous cosmic fluid takes the form
\begin{equation}
T_{\mu\nu} = (\rho + \bar{p})u_\mu u_\nu - \bar{p} g_{\mu\nu},
\label{E10}
\end{equation}
where $\rho$ represents the matter-energy density, $u^\mu$ is the four-velocity components, and $u^\mu u_\mu = 1$.
For the choice of the matter Lagrangian density $\mathcal{L}_m=-\bar p$, which depends only on the matter variables and not explicitly on the metric tensor, the second-order derivative term in Eq.~(\ref{E2.6}) vanishes,
\begin{equation}
\frac{\partial^2\mathcal{L}_m}{\partial g^{\mu\nu}\partial g^{\alpha\beta}}=0,
\label{E11}
\end{equation}
consequently, Eq.~(\ref{E2.6}) reduces to
\begin{equation}
\Theta_{\mu\nu} = -2T_{\mu\nu} - \bar{p} g_{\mu\nu}.
\label{E12}
\end{equation}
The thermodynamic pressure of the cosmic fluid is assumed to satisfy the barotropic equation of state
\begin{equation}
p=w\rho,
\label{E13}
\end{equation}
where $w$ is a constant EoS parameter. Throughout this work, we restrict the analysis to the non-phantom regime by adopting the prior $-1\leq w\leq1$. This interval encompasses the main physically relevant cosmological fluids, including vacuum energy ($w=-1$), pressureless matter ($w=0$), and radiation ($w=1/3$). Following Ref.~\cite{Ren2006}, the effective pressure of the bulk viscous fluid is therefore written as
\begin{equation}
\bar{p}=w\rho-3\xi H.
\label{E2.14}
\end{equation}
Substituting the above relations into the modified field equations leads to the final expression governing the cosmological dynamics in the presence of bulk viscosity,
\begin{equation}
R_{\mu\nu} - \frac{1}{2}R g_{\mu\nu}
= T_{\mu\nu} + 2f_T T_{\mu\nu} + \left[2\bar{p}f_T + f(T)\right] g_{\mu\nu}.
\label{E15}
\end{equation}

\section{Cosmological Solutions}
\label{sec3}
To study the cosmological evolution of the proposed viscous models, we consider a homogeneous and isotropic space-time, described by the FLRW line element \cite{Koussour2024}
\begin{equation}
ds^2 = dt^2 - a^2(t)\left[dr^2 + r^2(d\theta^2 + \sin^2\theta d\phi^2)\right],
\label{E16}
\end{equation}
where $a(t)$ denotes the cosmic scale factor. For this geometry, the Ricci scalar takes the form
\begin{equation}
R = -6(\dot{H} + 2H^2),
\label{E17}
\end{equation}
where the Hubble parameter is defined by $H=\frac{\dot a}{a}$, which characterizes the expansion rate of the Universe.
Within the framework of viscous cosmology, we adopt the linear function $f(T)=\lambda T$ \cite{Harko2011,Moraes2016,Myrzakulov2012}, where $\lambda$ is a constant coupling parameter describing the interaction between matter and geometry. Under this assumption, the modified Friedmann equations of $f(R,T)$ gravity become
\begin{equation}
3H^2 = (1 + 3\lambda)\rho - \lambda \bar{p}.
\label{E3.3}
\end{equation}
\begin{equation}
2\dot{H} + 3H^2 = \lambda \rho - (1 + 3\lambda)\bar{p}.
\label{E3.4}
\end{equation}
It is important to mention that the standard viscous cosmology is recovered in the limit $\lambda=0$. Combining Eqs.~(\ref{E3.3}) and~(\ref{E3.4}), one can derive the matter-energy density in compact form
\begin{equation}
\rho = \frac{(3 + 6\lambda)H^2 - 2\lambda \dot{H}}{(1 + 3\lambda)^2 - \lambda^2}.
\label{E20}
\end{equation}

\subsection{Model A: Dust bulk viscous}
For the dust fluid ($w=0$), the effective pressure, Eq.~(\ref{E2.14}), becomes
\begin{equation}
\bar p=-3\xi H,
\label{E21}
\end{equation}
where $\xi$ denotes the bulk viscosity coefficient.
The evolution equation of the Hubble parameter is then obtained, from Eqs.~(\ref{E3.3}) and~(\ref{E3.4}), as

\begin{equation}
\dot H+
\frac{3(2\lambda+1)}
{2\left(1+3\lambda\right)}
H^{2}
-
\frac{3\xi(2\lambda+1)(4\lambda+1)}
{2(1+3\lambda)}
H=0.
\label{E3.7}
\end{equation}
Or in term of the redshift, $z=\frac{a_0}{a-1}$, Eq.~(\ref{E3.7}) becomes
\begin{equation}
(1+z)\frac{dH}{dz}
-\frac{3}{2}\frac{(2\lambda+1)}{1+3\lambda}H
+\frac{3}{2}\frac{\xi(2\lambda+1)(4\lambda+1)}{1+3\lambda}
=0.
\label{E23}
\end{equation}
Finally, by integrating the resulting equation, we obtain the Friedmann equation,
\begin{equation}
E(z)=(1+z)^{\frac{3(2\lambda+1)}
{2(1+3\lambda)}}
+\frac{\xi}{H_{0}}(4\lambda+1)
[1-(1+z)^{\frac{3(2\lambda+1)}
{2(1+3\lambda)}}
],
\label{E24}
\end{equation}
with $E(z)=H/H_{0}$ is the dimensionless Hubble rate and ${H_{0}}$ is the present value of the Hubble parameter.

\subsection{Model B: General bulk viscous}

We now generalize the previous analysis by allowing the EoS parameter $w$ to vary freely.
The effective pressure becomes
\begin{equation}
\bar p=w\rho-3\xi H.
\label{25}
\end{equation}
Substituting the effective pressure into the field equations yields

\begin{equation}
\dot H
+
\frac{3(1+w)(2\lambda+1)}
{2\left[1+\lambda(3-w)\right]}
H^{2}
-
\frac{3\xi(2\lambda+1)(4\lambda+1)}
{2\left[1+\lambda(3-w)\right]}
H
=0.
\label{E26}
\end{equation}
After changing the variable from cosmic time to redshift, the Friedmann equation is obtained as
\begin{equation}
E(z)
=
(1+z)^{
\frac{3(1+w)(2\lambda+1)}
{2[1+\lambda(3-w)]}
}
+
\frac{\xi}{H_{0}}
\left(\frac{4\lambda+1}{1+w}\right)
\left[
1-
(1+z)^{
\frac{3(1+w)(2\lambda+1)}
{2[1+\lambda(3-w)]}
}
\right].
\label{E27}
\end{equation}

\section{Observational Data}
\label{sec4}
To constrain the free parameters of the proposed cosmological models, we make use of several recent cosmological observations. Our analysis is based on the DESI-DR2 measurements \cite{DESI2025II,DESI2025I}, Pantheon$^+$+SH0ES \cite{PantheonPlus2022,Scolnic2022,Brout2022}, and $H(z)$ measurements \cite{Jimenez2002,Moresco2012,Moresco2016}.
The estimation of the model parameters is performed through the MCMC technique \cite{Padilla2021}, which samples the posterior probability distributions by efficiently exploring the multidimensional parameter space.

For the standard $\Lambda$CDM model, the parameter vector is chosen as $ \Theta_{\Lambda \mathrm{CDM}}=\left\{M_B,\,\Omega_m,\,r_d ,\,h\right\}$, where $M_B$ denotes the absolute magnitude of Type Ia supernovae, $\Omega_m$ is the present matter density parameter, $r_d$ represents the sound horizon at the drag epoch, and $h=H_0/100$ is the reduced Hubble constant.

For model A, the parameter space is extended to include the matter--geometry coupling parameter $\lambda$ together with the bulk viscosity coefficient $\xi$, namely $\Theta_{\mathrm{A}}=\left\{M_B,\,r_d,\,\,\lambda,\,h,\,\xi\right\}$.
In model B, the EoS parameter $w$ is treated as an additional free parameter. The corresponding parameter vector therefore becomes $\Theta_{\mathrm{B}}=\left\{M_B,\,r_d,\,\lambda,\,h,\,w,\,\xi\right\}$.
The prior intervals adopted for all cosmological parameters are summarized in Table~\ref{tab:1}.
\begin{table*}[ht]
\centering
\caption{Prior imposed on different parameters for A, B and $\Lambda$CDM models}
\label{tab:1}
\begin{tabular}{lc}
\hline
Parameters              &Prior  \\
\hline
$M_B$                & [-20, -18] \\

$\Omega_m$                & [0.25, 0.33] \\

$r_d$ & [100, 200] \\

$\lambda$ & [-2, 2] \\

$h$ & [0.4, 1] \\

$\omega$ & [-1, 1] \\

$\xi$ & [0, 200] \\
\hline
\end{tabular}
\end{table*}
\subsection{DESI-DR2 }

The baryon acoustic oscillation (BAO) measurements from the DESI-DR2 survey provide powerful geometrical constraints on the expansion history of the Universe over the redshift range $0.295 \leq z \leq 2.33$    \cite{AbdulKarim2025}. In this work, we use the DESI-DR2 BAO measurements (denoted as DESI-DR2 in the following) obtained from the BGS, LRG, ELG, QSO, and Ly$\alpha$ forest tracers. These measurements provide constraints on the comoving angular diameter distance $D_M(z)$, the Hubble distance $D_H(z)$, and the volume-averaged distance $D_V(z)$, expressed in units of the sound horizon at the drag epoch $r_d$ \cite{Eisenstein1998}. The corresponding effective redshifts and BAO observables are summarized in Table IV of Ref.~\cite{AbdulKarim2025}.

The residual vector is defined as
\begin{equation}
\Delta\vec{V}
=
\vec{V}_{\rm obs}
-
\vec{V}_{\rm th},
\label{E28}
\end{equation}
where $\vec{V}_{\rm obs}$ and $\vec{V}_{\rm th}$ denote the vectors of the observed and theoretical BAO quantities, respectively, and explicitly,

\begin{equation}
\vec{V}
=
\left(
\frac{D_M(z)}{r_d},
\frac{D_H(z)}{r_d},
\frac{D_V(z)}{r_d}
\right)^T,
\label{E29}
\end{equation}
with the comoving angular diameter distace, $D_M(z)$, the Hubble distance, $D_H(z)$, and the volume-averaged distance, $D_V(z)$, are given by

\begin{equation}
D_M(z)=c\int_{0}^{z}\frac{dz'}{H(z')},
\label{E30}
\end{equation}

\begin{equation}
D_H(z)=\frac{c}{H(z)},
\label{E31}
\end{equation}
and
\begin{equation}
D_V(z)=\left[zD_H(z)D_M^2(z)\right]^{1/3},
\label{E32}
\end{equation}
respectively. Here, $c$ denotes the speed of light, $H(z)$ is the Hubble expansion rate predicted by the cosmological model.

The corresponding chi-square statistic is

\begin{equation}
\chi^2_{\rm DESI-DR2}
=
\Delta\vec{V}^{T}
C^{-1}_{\rm DESI-DR2}
\Delta\vec{V},
\label{E33}
\end{equation}
where $C_{\rm DESI-DR2}$ is the covariance matrix associated with the DESI-DR2 measurements.

\subsection{Pantheon$^+$+SH0ES}

The luminosity-distance information from Type Ia supernovae is incorporated using the Pantheon+ compilation, which consists of 1701 light curves of 1550 spectroscopically confirmed SNIa covering the redshift range $0.001 \leq z \leq 2.26$ \cite{Brout2022,Scolnic2022}. Owing to its precision and broad redshift coverage, this dataset provides one of the strongest observational probes of the late-time expansion history.

The corresponding likelihood is constructed from the covariance matrix of the observed distance modulus and is expressed through the chi-square function 

\begin{equation}
\chi^2_{\rm Pantheon+}
=
\Delta\vec{\mu}^{T}
C^{-1}_{\rm Pantheon+}
\Delta\vec{\mu},
\label{E4.7}
\end{equation}
where $C_{\rm Pantheon+}$ denotes the full covariance matrix supplied by the Pantheon+ measurements, the residual vector is defined as

\begin{equation}
\Delta\vec\mu_i
=
\mu_i^{\rm obs}
-
\mu^{\rm th}(z_i),
\label{E35}
\end{equation}
where $\mu_i^{\rm obs}$ is the observed distance modulus, while $\mu^{\rm th}(z_i)$ is the theoretical distance modulus given by,

\begin{equation}
\mu^{\rm th}(z)
=
5\log_{10}
\left(
\frac{d_L(z)}{\rm Mpc}
\right)
+25,
\label{E36}
\end{equation}
where the luminosity distance, $d_L(z)$, is calculated as

\begin{equation}
d_L(z)
=
(1+z)c
\int_0^z
\frac{dz'}{H(z')},
\label{E37}
\end{equation}
with $c$ is the speed of light.

While the standard Pantheon+ likelihood is formulated in terms of the distance modulus residuals, the Pantheon$^+$+SH0ES analysis adopts a modified residual vector that explicitly accounts for the Cepheid calibration of nearby host galaxies.

For the Pantheon$^+$+SH0ES analysis (denoted as PPS in the following), a subset of Type Ia supernovae is calibrated using Cepheid distances from the SH0ES program \cite{Riess2022}, allowing the absolute magnitude $M_B$ to be directly constrained \cite{Brout2022}. Consequently, the residual vector is modified according to

\begin{equation}
\vec F_i=
\begin{cases}
m_{B,i}^{\rm obs}-M_B-\mu_i^{\rm Cepheid},
&
i\in {\rm Cepheid\ hosts},
\\[2mm]
m_{B,i}^{\rm obs}
-
M_B
-
\mu_{\rm model}(z_i),
&
{\rm otherwise},
\end{cases}
\label{E38}
\end{equation}
where $m_{B,i}^{\rm obs}$ denotes the corrected apparent peak magnitude, $\mu_i^{\rm Cepheid}$ is the Cepheid distance modulus provided by the SH0ES-calibrated subsample, and $\mu_{\rm model}(z_i)$ is the corresponding theoretical distance modulus.

Accordingly, the equation~(\ref{E4.7}) becomes

\begin{equation}
\chi^2_{\rm PPS}
=
\vec{F}^ {T}
C^{-1}_{\rm PPS}
\vec{F}.
\label{E39}
\end{equation}
with $C_{\rm PPS}$ is the full covariance matrix of Pantheon$^+$+SH0ES \cite{Scolnic2022}.

\subsection{$H(z)$ measurements}

To constrain the cosmological parameters, we make use of a compilation of 36 independent measurements of the Hubble expansion rate covering the redshift interval $0.07 \leq z \leq 2.34$. The dataset consists of 30 cosmic chronometer (CC) measurements derived from the differential ages of passively evolving galaxies and 6 additional measurements obtained from radial baryon acoustic oscillations (BAO) and other observational techniques \cite{Jimenez2002,Moresco2012,Moresco2016}.

The Hubble expansion rate is determined from the differential evolution of the redshift according to

\begin{equation}
H(z)
=
-\frac{1}{1+z}\frac{dz}{dt},
\label{E40}
\end{equation}
where $z$ is the cosmological redshift and $t$ denotes the cosmic time.

For the statistical analysis, the likelihood of the $H(z)$ dataset (denoted as H in the following) is quantified through the chi-square estimator

\begin{equation}
\chi^2_{H}
=
\sum_{i=1}^{N_H}
\frac{\left[H^{\rm obs}(z_i)-H^{\rm th}(z_i,\Theta)\right]^2}
{\sigma_{H,i}^{2}},
\label{E41}
\end{equation}
where $N_H=36$ is the total number of Hubble parameter measurements, $H^{\rm obs}(z_i)$ represents the observed value at redshift $z_i$, while $H^{\rm th}(z_i,\Theta)$ denotes the corresponding theoretical prediction of the cosmological model characterized by the parameter set $\Theta$. The quantity $\sigma_{H,i}$ is the corresponding measurement uncertainty.

\subsection{Total chi-square}

To obtain robust constraints on the free parameters of the proposed cosmological models, we perform a joint analysis by combining all the observational datasets considered in this work. Assuming that these datasets are statistically independent, the total chi-square function is constructed as the sum of the individual chi-square contributions,

\begin{equation}
\chi^2_{\rm tot}
=
\chi^2_{\rm DESI-DR2}
+
\chi^2_{\rm PPS}
+
\chi^2_{H},
\label{E42}
\end{equation}
where $\chi^2_{\rm DESI-DR2}$, $\chi^2_{\rm PPS}$, and $\chi^2_{H}$ represent the chi-square functions associated with DESI-DR2+ PPS+ H datasets, respectively. The best-fit values of the model parameters are determined by minimizing the total chi-square function.

\subsection{Information criteria}

In addition to estimating the cosmological parameters, it is essential to evaluate the statistical quality of the proposed models while taking into account the number of adjustable parameters. Since adding more free parameters generally reduces the minimum chi-square value, a model with the lowest $\chi^2_{\min}$ is not necessarily the most appropriate description of the observational data. For this purpose, we employ the Akaike Information Criterion (AIC) \cite{Akaike1974}, the corrected Akaike Information Criterion (AICc) \cite{Burnham2002}, and the Bayesian Information Criterion (BIC)\cite{Schwarz1978}, which allow a balanced comparison between our viscous cosmological models and the standard $\Lambda$CDM scenario.

The Akaike Information Criterion is defined as
\begin{equation}
AIC=\chi^2_{\min}+2K
\label{E43}
\end{equation}
where $\chi^2_{\min}$ denotes the minimum of the chi-square value obtained from the fit, while $K$ represents the total number of free parameters of the model.
For completeness, the corrected Akaike Information Criterion (AICc) is defined as \cite{Burnham2002}
\begin{equation}
AICc=AIC+\frac{2K(K+1)}{N-K-1},
\label{E44}
\end{equation}
where $N$ denotes the total number of observational data points.
To compare each model with the reference $\Lambda$CDM model, we evaluate

\begin{equation}
\Delta AICc=AICc_{\mathrm{model}}-AICc_{\Lambda\mathrm{CDM}}.
\label{E45}
\end{equation}
The preferred model is the one with the lowest AICc value. The interpretation of $\Delta$AICc follows the standard statistical convention. Models with $|\Delta AICc|<2$ are considered to have support comparable to that of the reference model. Values in the range $2\leq|\Delta AICc|<4$ indicate moderate evidence against the model with the larger information criterion, whereas $4\leq|\Delta AICc|<10$ corresponds to considerably weaker support. When $|\Delta AICc|\geq10$, the observational data strongly favor the model with the lower information criterion. To complement this analysis, we also compute the Bayesian Information Criterion (BIC), defined as
\begin{equation}
BIC=\chi^2_{\min}+K\ln N.
\label{E46}
\end{equation}

The relative BIC is computed as
\begin{equation}
\Delta BIC=BIC_{\mathrm{model}}-BIC_{\Lambda\mathrm{CDM}}.
\label{E47}
\end{equation}

Compared with the AIC, BIC introduces a stronger penalty for models containing additional free parameters. Following the commonly adopted interpretation, $|\Delta BIC|<2$ indicates that the evidence against the model with the higher BIC values is weak, $2\leq|\Delta BIC|<6$ corresponds to positive evidence, $6\leq |\Delta BIC|<10$ indicates strong evidence, whereas $|\Delta BIC|\geq10$ is generally interpreted as very strong support in favor of the model with the lower BIC values. Throughout this work, the standard $\Lambda$CDM model is adopted as the reference scenario. Consequently, the AICc and BIC statistics are calculated for models A and B using the combined DESI-DR2, Pantheon +, and H(z) datasets. These statistical indicators provide an objective assessment of whether the additional parameters introduced by the proposed viscous cosmological models are justified by the improvement in the agreement with the observational data.

\section{Results and Discussion}
\label{sec5}
\begin{figure}[H]
    \centering
    \includegraphics[width=0.6\textwidth]{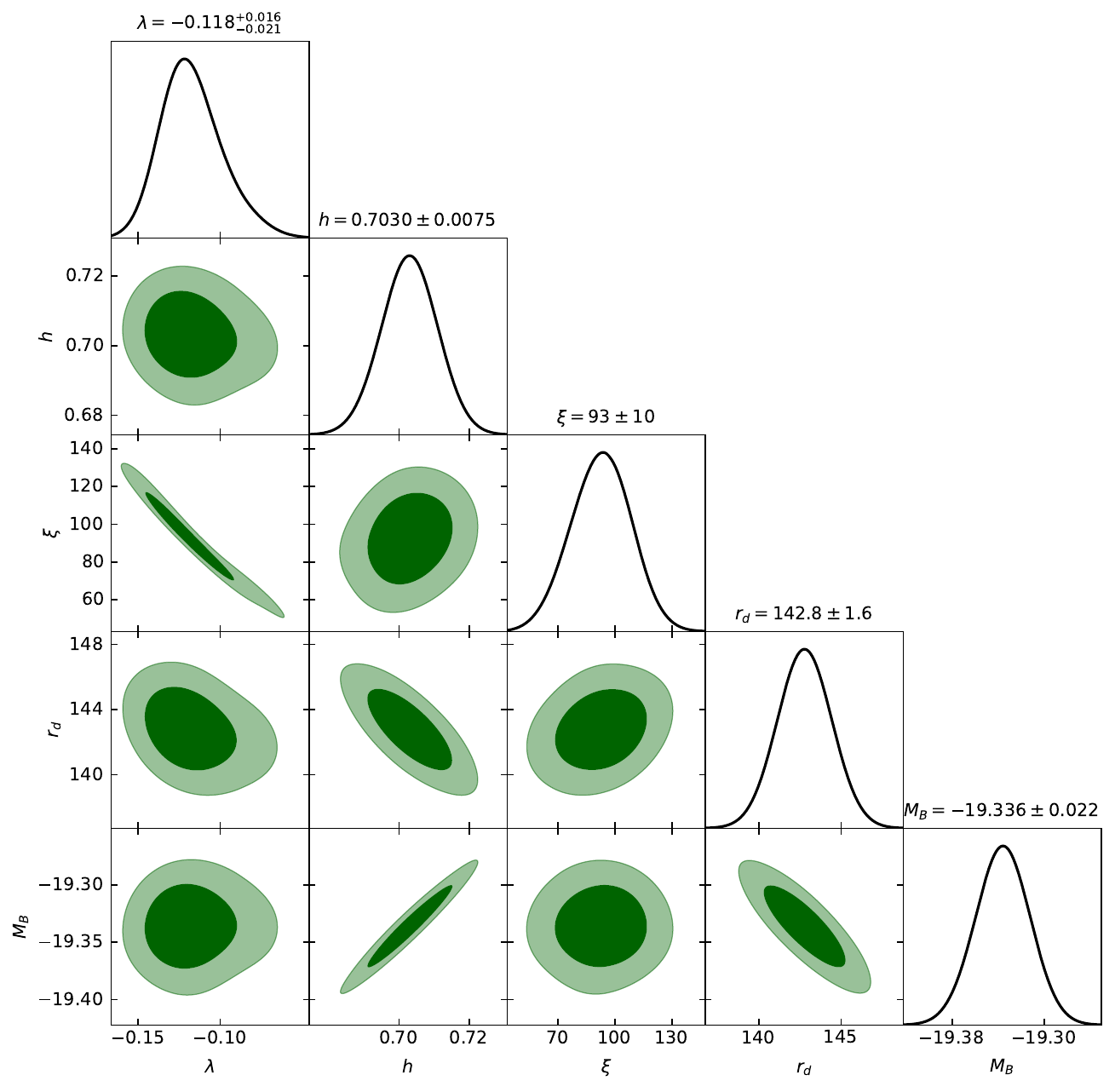}
    \caption{The confidence contours at $1\sigma$ and $2\sigma$ and the $1D$ posterior distributions obtained from DESI-DR2+ PPS+ H datasets for model A.}
    \label{fig:contours A}
\end{figure}

\begin{figure}[H]
    \centering
    \includegraphics[width=0.6\textwidth]{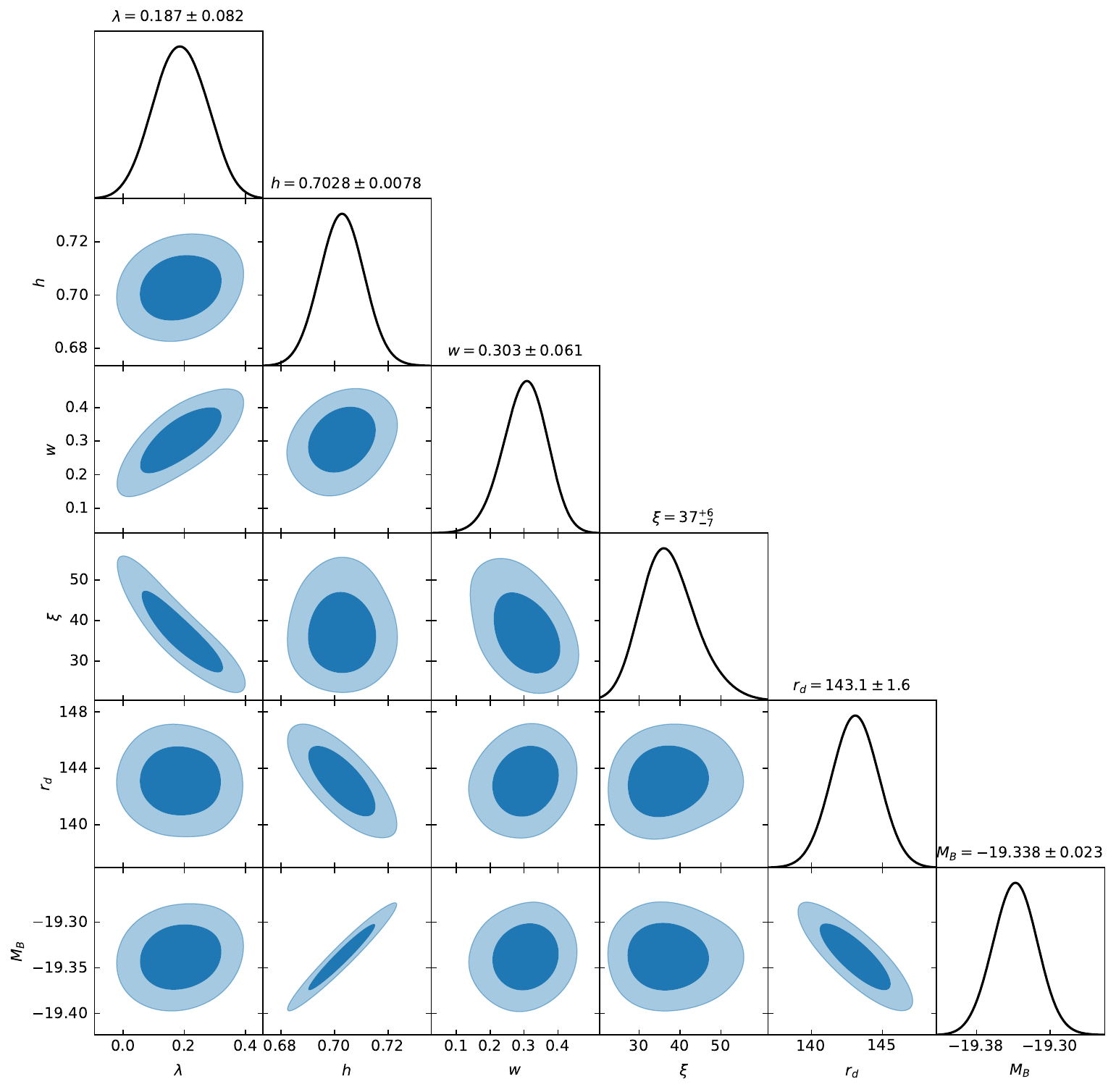}
    \caption{The confidence contours at $1\sigma$ and $2\sigma$ and the $1D$ posterior distributions obtained from DESI-DR2+ PPS+ H datasets for model B.}
    \label{fig:contours B}
\end{figure}

Table~\ref{tab:results} summarizes the mean values of the cosmological parameters together with their corresponding $1\sigma$ uncertainties for $\Lambda$CDM, model A and model B obtained from DESI-DR2+ PPS+ H datasets. Table~\ref{tab:results} also lists the minimum of the chi-square values together with the corresponding $\Delta$AICc, and $\Delta$BIC statistics. Figs.~(\ref{fig:contours A}) and~(\ref{fig:contours B}) present the two-dimensional confidence contours and the marginalized one-dimensional posterior distributions for the free parameters of models A and B, respectively.
\begin{table*}[ht]
\centering
\caption{The mean$\pm1\sigma$ values of the cosmological parameters and statistical results obtained from DESI-DR2+ PPS+ H datasets.}
\label{tab:results}
\renewcommand{\arraystretch}{1.2}
\begin{tabular}{lccc}
\toprule
\multicolumn{4}{c}{\textbf{ DESI-DR2+ PPS+ H }}\\
\hline
\textbf{Parameter} & $\Lambda$CDM & Model A & Model B\\
\hline
$M_B$ & $-19.329\pm0.022$ & $-19.336\pm0.022$ & $-19.338\pm0.023$\\

$\Omega_m$ & $0.2910\pm0.0064$ & --- & ---\\

$r_d$ & $143.6\pm1.5$ & $142.8\pm1.6$ & $143.1\pm1.6$\\

$\lambda$ & --- & $-0.118^{+0.016}_{-0.021}$ & $0.187\pm0.082$\\

$h$ & $0.7119\pm0.0074$ & $0.7030\pm0.0075$ & $0.7028\pm0.0078$\\

$\omega$ & --- & $0$ & $0.303\pm0.061$\\

$\xi$ & --- & $93\pm10$ & $37^{+6}_{-7}$\\

\hline
\multicolumn{4}{c}{\textbf{Statistical results}}\\
\hline

$\chi^2_{\rm min}$ & 1575.29 & 1571.15 & 1571.14\\

$\Delta$AICc & 0 & -2.13 & -0.12\\

$\Delta$BIC & 0 & 3.32 & 10.78\\
\hline
\end{tabular}
\end{table*}

As shown in Table~\ref{tab:results}, DESI-DR2+ PPS+ H datasets provides well-constrained cosmological parameters for the three cosmological scenarios. The marginalized one-dimensional posterior distributions exhibit nearly Gaussian shapes, while the corresponding two-dimensional confidence contours remain compact, indicating well-constrained cosmological parameters. For model A, a strong anticorrelation is observed between $\lambda$ and $\xi$, while a pronounced positive correlation is found between $h$ and $M_B$, together with noticeable anticorrelations between $h$ and $r_d$ and between $r_d$ and $M_B$. For model B, the contours reveal a strong positive correlation between $\omega$ and $\lambda$, as well as a strong anticorrelation between $\lambda$ and $\xi$. Other parameter pairs exhibit weaker correlations. Overall, the compact contours and well-behaved posterior distributions indicate that the combined observational datasets provide stable and meaningful constraints on the free parameters of $\Lambda$CDM, model A, and model B scenarios.

For the standard $\Lambda$CDM model, the best-fit parameters are found to be $M_B=-19.329\pm0.022$, $r_d=143.6\pm1.5$ Mpc, and $h=0.7119\pm0.0074$. For model A, corresponding to the dust fluid case ($w=0$), the matter--geometry coupling parameter is constrained to $\lambda=-0.118^{+0.016}_{-0.021}$, indicating that DESI-DR2+ PPS+ H datasets favor a weak negative coupling between matter and geometry. The Hubble constant is determined as $h=0.7030\pm0.0075$, while the bulk viscosity coefficient is found to be $\xi=93\pm10$, confirming that a non-vanishing viscous contribution is compatible with DESI-DR2+ PPS+ H datasets and may play an important role in the late-time cosmic evolution. Model B extends the previous scenario by treating the EoS parameter as a free parameter. The MCMC analysis yields $w=0.303\pm0.061$, together with a positive matter--geometry coupling, $\lambda=0.187\pm0.082$, and a reduced bulk viscosity coefficient, $\xi=37^{+6}_{-7}$. Compared with model A, the decrease in the viscosity coefficient suggests that part of the accelerated expansion can be effectively described by the additional pressure associated with the free EoS parameter.

The statistical performance of the three cosmological models is evaluated using the minimum chi-square together with AICc, and BIC. Although the $\Lambda$CDM model provides an excellent fit to DESI-DR2+ PPS+ H datasets, both modified gravity models achieve slightly lower minimum chi-square values, namely $\chi^2_{\rm min}=1571.15$ for model A and $\chi^2_{\rm min}=1571.14$ for model B.
Owing to this improvement, model A yields $\Delta{\rm AICc}=-2.13$, indicating that the better quality of the fit compensates for the introduction of one additional free parameter. By contrast, BIC gives $\Delta{\rm BIC}=3.32$, showing that the modest improvement in the goodness of fit is insufficient to offset the stronger complexity penalty imposed by the additional parameter. Consequently, while model A is slightly preferred over the standard $\Lambda$CDM model according to AICc, the BIC continues to favor the simpler $\Lambda$CDM scenario.
For model B, the minimum chi-square is only marginally smaller than that of model A; however, this negligible improvement does not justify the introduction of an additional free parameter associated with the EoS parameter. As a result, the model gives $\Delta{\rm AICc}=-0.12$, indicating a statistical performance nearly indistinguishable from that of $\Lambda$CDM according to AICc. Nevertheless, BIC yields $\Delta{\rm BIC}=10.78$, providing strong evidence in favor of the standard cosmological model. Therefore, allowing the EoS parameter to vary freely does not produce a statistically significant improvement over model A.
Overall, the information criteria reveal a consistent picture. The AICc results, which emphasizes predictive performance, shows a slight preference for model A owing to its improved fit, whereas the BIC, which imposes a stronger penalty on model complexity for large datasets, favors the minimal $\Lambda$CDM model. These results indicate that the inclusion of matter--geometry coupling and bulk viscosity can improve the description of late-time cosmological observations without dramatically increasing model complexity, although the statistical significance of this improvement depends on the adopted model-selection criterion. Our findings are also consistent with recent observational studies of viscous cosmology in $f(R,T)$ gravity, confirming that matter--geometry coupling together with bulk viscosity constitutes a viable framework for describing the late-time Universe \cite{Koussour2024,Bouali2023,Errahmani2026}.

\begin{figure}[H]
    \centering
    \includegraphics[width=0.75\linewidth]{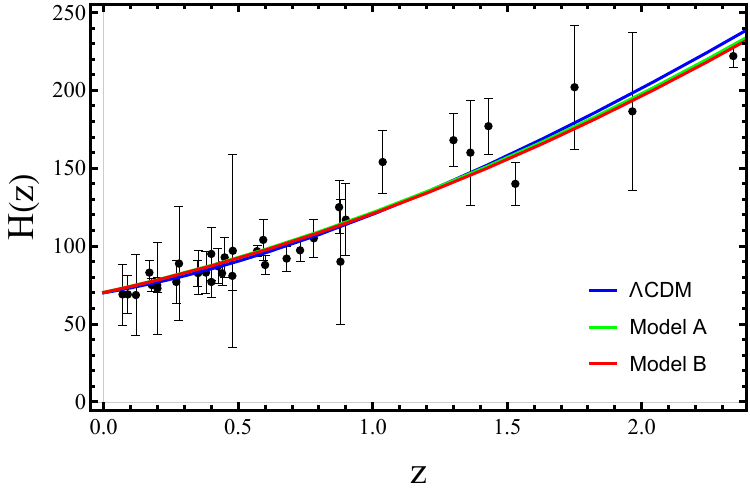}
    \caption{Evolution of the Hubble parameter $H(z)$ as a function of redshift for $\Lambda$CDM, model A and model B.}
     \label{fig:Hz}
\end{figure}

\begin{figure}[H]
    \centering
    \includegraphics[width=0.75\linewidth]{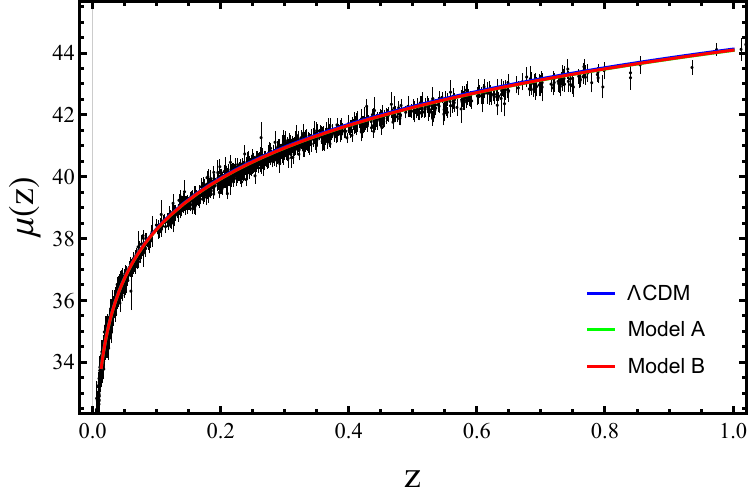}
    \caption{Comparison between the Pantheon+ Type Ia supernova data \cite{Scolnic2022} and the theoretical distance modulus $\mu(z)$ predicted by $\Lambda$CDM, model A and model B.}
    \label{fig:muz}
\end{figure}

\begin{figure}[H]
    \centering
    \includegraphics[width=0.75\linewidth]{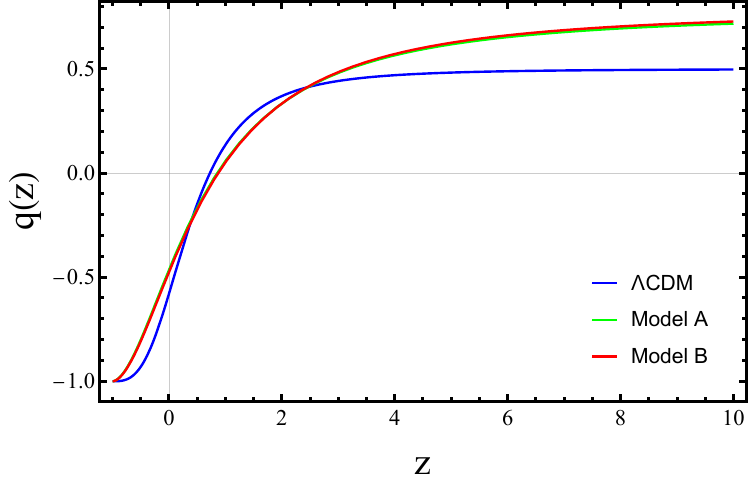}
    \caption{Evolution of $q(z)$ as a function of redshift, for models A, B, and $\Lambda$CDM.}
    \label{fig:qz}
\end{figure}

\begin{figure}[H]
    \centering
    \includegraphics[width=0.7\linewidth]{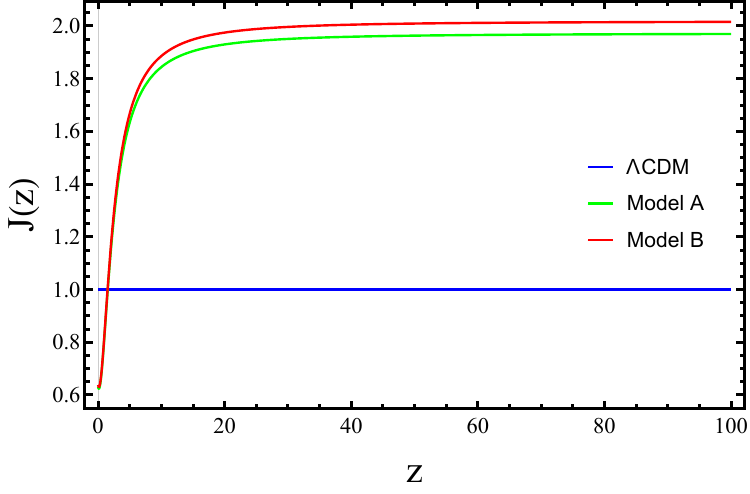}
    \caption{Evolution of the jerk parameter $j(z)$ with respect to redshift z, for models A, B, and $\Lambda$CDM.}
    \label{fig:jz}
\end{figure}

\begin{figure}[H]
    \centering
    \includegraphics[width=0.7\linewidth]{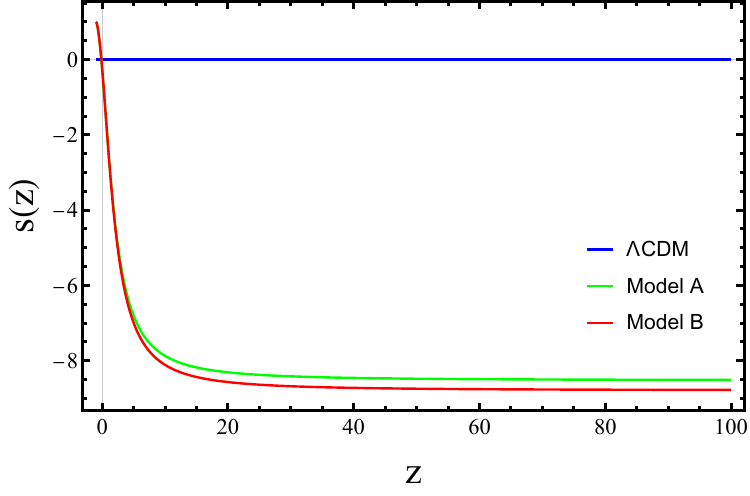}
    \caption{Evolution of the snap parameter $s(z)$ with respect to redshift, for models A, B, and $\Lambda$CDM.}
    \label{fig:sz}
\end{figure}

The cosmological evolution predicted by the three models is further illustrated in Figs.~(\ref{fig:Hz}--\ref{fig:sz}) through the Hubble parameter, the distance modulus, and the cosmographic parameters.
Fig.~(\ref{fig:Hz}) presents the evolution of the Hubble parameter $H(z)$ together with the observational Hubble data. The theoretical predictions of the $\Lambda$CDM model, model A, and model B all provide an excellent description of the observed expansion history over the considered redshift range. Although the three curves are nearly indistinguishable at low redshifts, small deviations become visible at intermediate and high redshifts due to the effects of the matter--geometry coupling and bulk viscosity introduced in the modified gravity models. Interestingly, the slightly lower minimum $\chi^2$ values obtained for models A and B indicate that these additional physical effects improve the agreement with the observational Hubble measurements without producing significant departures from the standard cosmological evolution.

A similar behavior is observed in Fig.~(\ref{fig:muz}), where the theoretical distance modulus is compared with the Pantheon+ Type Ia supernova sample. The predicted distance modulus of the three cosmological scenarios almost overlap over the entire redshift interval, demonstrating that all models successfully reproduce the luminosity distance relation inferred from supernova observations. The excellent agreement between the theoretical curves and the Pantheon+ data confirms that the proposed $f(R,T)$ viscous cosmologies preserve the successful late-time expansion history described by the standard $\Lambda$CDM model while allowing small observationally consistent deviations associated with the matter--geometry coupling and viscous effects.

To further investigate the dynamical evolution of the proposed cosmological models, we consider the cosmographic parameters, namely the deceleration $q(z)$, jerk $j(z)$, and snap $s(z)$ parameters\cite{Visser2004,Cattoen2007,Dunsby2016}. Expressed as functions of the redshift $z$, these quantities can be written as
\begin{equation}
q(z)=-1+\frac{(1+z)}{H(z)}\frac{dH(z)}{dz},
\label{E48}
\end{equation}
\begin{equation}
j(z)=\frac{(1+z)^2H''(z)}{H(z)}+q^2(z),
\label{E49}
\end{equation}
and
\begin{equation}
s(z)=-(1+z)\frac{dj(z)}{dz}-j(z)\left[2+3q(z)\right],
\label{E50}
\end{equation}
respectively, where $H(z)$ denotes the Hubble parameter and the derivatives are taken with respect to the redshift. These cosmographic quantities provide complementary information about the expansion history of the Universe and are particularly useful for distinguishing different cosmological models beyond the standard $\Lambda$CDM scenario\cite{Capozziello2011,Aviles2012,Dunsby2016}.

Furthermore, the cosmographic parameters provide additional insight into the dynamical evolution of the Universe. Fig.~(\ref{fig:qz}) illustrates the evolution of the deceleration parameter $q(z)$ for the $\Lambda$CDM model together with models A and B. At high redshifts, all three models predict positive values of $q(z)$, corresponding to the matter-dominated decelerating phase of the Universe. As the redshift decreases, the deceleration parameter gradually becomes negative, indicating the transition to the present accelerated expansion. The transition redshift is found to be $z_t\simeq0.726$ for the $\Lambda$CDM model, while models A and B predict $z_t\simeq0.855$ and $z_t\simeq0.872$, respectively. Therefore, both viscous $f(R,T)$ models undergo the transition to the accelerated expansion of the Universe at an earlier epoch than the standard $\Lambda$CDM scenario.
Although the overall behavior remains very similar, slight quantitative differences are observed between the modified gravity models and the standard $\Lambda$CDM scenario. These deviations originate from the combined effects of the matter--geometry coupling and the bulk viscous pressure, which modify the effective expansion dynamics while remaining consistent with the observationally inferred cosmic expansion history. The close agreement of the three curves at both low and high redshifts further demonstrates that the proposed viscous $f(R,T)$ models successfully reproduce the observed transition from decelerated to accelerated expansion while introducing only moderate deviations from the standard cosmological model.

Fig.~(\ref{fig:jz}) displays the evolution of the jerk parameter $j(z)$ for the $\Lambda$CDM model together with models A and B. As expected, the standard $\Lambda$CDM scenario predicts a constant value $j=1$ over the entire redshift interval, which constitutes one of its distinctive characteristics. In contrast, the two viscous $f(R,T)$ cosmological models exhibit a dynamical evolution of the jerk parameter. 
Starting from values below unity at low redshifts, both models increase rapidly before approaching nearly constant asymptotic values at higher redshifts. Throughout the considered range, model~B predicts slightly larger values of $j(z)$ than model~A, although the difference between the two models remains relatively small. These deviations from the constant $\Lambda$CDM prediction arise from the combined effects of the matter--geometry coupling and bulk viscosity, which modify the higher-order dynamics of the cosmic expansion. Nevertheless, the smooth evolution of the jerk parameter indicates that both modified gravity models remain physically well behaved while extending the standard cosmological scenario.

Finally, Fig.~(\ref{fig:sz}) illustrates the evolution of the snap parameter $s(z)$ for the $\Lambda$CDM model, model~A, and model~B. While the standard $\Lambda$CDM model predicts a constant value close to zero throughout the cosmic evolution, the viscous $f(R,T)$ models exhibit a markedly different behavior. Both model~A and model~B start from positive values at very low redshifts, then decrease rapidly and become negative before gradually approaching nearly constant values at high redshifts. The evolution of the snap parameter therefore reveals the impact of the matter--geometry coupling of the cosmic expansion. Moreover, model~B consistently predicts slightly lower values of $s(z)$ than model~A, reflecting the additional degree of freedom associated with the EoS parameter. Despite these quantitative differences, both models display smooth and stable evolutions, confirming that the inclusion of bulk viscosity and matter--geometry coupling modifies the expansion dynamics without introducing pathological behavior. Consequently, the snap parameter provides an additional diagnostic supporting the cosmological viability of the proposed viscous $f(R,T)$ models.

\section{Conclusion}
\label{sec6}
In this work, we have investigated the cosmological consequences of bulk viscosity within the framework of the linear $f(R,T)$ gravity model, $f(R,T)=R+2\lambda T$, by considering two cosmological scenarios. In model A, the cosmic fluid is assumed to be pressureless ($w=0$), whereas in model B the EoS parameter is treated as a free parameter. For both models, analytical expressions for the Hubble parameter were derived and confronted with the latest cosmological observations\cite{Planck2018,Brout2022,DESI2025}.
The model parameters were constrained through a Markov Chain Monte Carlo analysis using DESI-DR2+ PPS+ H datasets. The obtained results show that both viscous $f(R,T)$ cosmologies provide an excellent description of the late-time expansion history of the Universe. The matter--geometry coupling parameter and the bulk viscosity coefficient are found to be tightly constrained, indicating that the inclusion of viscous effects remains fully compatible with current observational data\cite{Planck2018,Brout2022,DESI2025}.

A comparison based on the minimum $\chi^2$, AICc, and BIC reveals that the statistical assessment depends on the adopted model-selection criterion. While model A provides the best compromise between goodness of fit and model complexity according to AICc, owing to its improved fit with only one additional free parameter, BIC continues to favor the simpler $\Lambda$CDM model because of its stronger penalty on model complexity for large datasets. Although model B yields the lowest minimum $\chi^2$, the improvement is too small to justify the introduction of an additional free parameter, making its overall statistical performance comparable to $\Lambda$CDM according to AICc but less favorable according to BIC. Overall, these results indicate that the dust-viscous scenario (model A) represents the most competitive extension of the standard cosmological model within the considered $f(R,T)$ framework, while BIC still supports the minimal $\Lambda$CDM description.

The cosmographic investigation further supports the viability of the proposed models. Both scenarios successfully reproduce the transition from an early decelerated expansion to the current accelerated phase while exhibiting only moderate deviations from the standard $\Lambda$CDM predictions in the jerk and snap parameters. These deviations originate from the combined effects of matter--geometry coupling and bulk viscosity and therefore provide additional signatures that could be tested with future observations.

Overall, the present study shows that bulk viscosity, together with the matter--geometry coupling inherent to $f(R,T)$ gravity, constitutes a viable extension of the standard cosmological model. Although the current observational data\cite{Planck2018,Brout2022,DESI2025} do not provide compelling evidence requiring a departure from $\Lambda$CDM, they allow the existence of viscous modified-gravity scenarios that remain statistically competitive while introducing a richer cosmological phenomenology. Future high-precision surveys, including forthcoming DESI releases and others, will significantly improve the observational constraints and may clarify whether these effects play a measurable role in the dynamics of the late-time Universe.


\begin{thebibliography}{99}
\bibitem{Riess1998} 
A.~G.~Riess et al., ``Observational Evidence from Supernovae for an Accelerating Universe and a Cosmological Constant,'' Astron. J. \textbf{116}, 1009 (1998).

\bibitem{Filippenko1998}
A.~V.~Filippenko, ``Optical Spectra of Supernovae,'' Phys. Rep. \textbf{307}, 31--44 (1998).

\bibitem{Perlmutter1999}
S.~Perlmutter et al., ``Measurements of $\Omega$ and $\Lambda$ from 42 High-Redshift Supernovae,'' Astrophys. J. \textbf{517}, 565--586 (1999).

\bibitem{Ade2016}
P.~A.~R.~Ade et al., ``Planck 2015 results. XIII. Cosmological parameters,'' Astron. Astrophys. \textbf{594}, A13 (2016).

\bibitem{eBOSS2021} 
eBOSS Collaboration, S. Alam et al., ``Completed SDSS-IV extended Baryon Oscillation Spectroscopic Survey: Cosmological implications from two decades of spectroscopic surveys,'' Phys. Rev. D \textbf{103}, 083533 (2021).

\bibitem{Zehavi2011}
I.~Zehavi et al., ``Galaxy Clustering in the Completed SDSS Redshift Survey: The Dependence on Color and Luminosity,'' Astrophys. J. \textbf{736}, 59 (2011).

\bibitem{Bartelmann2001}
M.~Bartelmann and P.~Schneider, ``Weak Gravitational Lensing,'' Phys. Rep. \textbf{340}, 291--472 (2001).

\bibitem{Eisenstein2005}
D.~J.~Eisenstein et al., ``Detection of the Baryon Acoustic Peak in the Large-Scale Correlation Function of SDSS Luminous Red Galaxies,'' Astrophys. J. \textbf{633}, 560--574 (2005).

\bibitem{Blake2012}
C.~Blake et al., ``The WiggleZ Dark Energy Survey: Joint measurements of the expansion and growth history at $z<1$,'' Mon. Not. R. Astron. Soc. \textbf{425}, 405--414 (2012).

\bibitem{Parkinson2012}
D.~Parkinson et al., ``The WiggleZ Dark Energy Survey: Final data release and cosmological results,'' Phys. Rev. D \textbf{86}, 103518 (2012).

\bibitem{Kazin2014}
E.~A.~Kazin et al., ``The WiggleZ Dark Energy Survey: Improved distance measurements to $z=1$ with reconstruction of the baryon acoustic feature,'' Mon. Not. R. Astron. Soc. \textbf{441}, 3524--3542 (2014).

\bibitem{Alam2015}
S.~Alam et al., ``The Eleventh and Twelfth Data Releases of the Sloan Digital Sky Survey: Final Data from SDSS-III,'' Astrophys. J. Suppl. Ser. \textbf{219}, 12 (2015).

\bibitem{Beutler2016}
F.~Beutler et al., ``The clustering of galaxies in the completed SDSS-III Baryon Oscillation Spectroscopic Survey: Baryon Acoustic Oscillations in Fourier space,'' Mon. Not. R. Astron. Soc. \textbf{455}, 3230--3248 (2016).

\bibitem{Planck2018} 
Planck Collaboration, N.~Aghanim et al., ``Planck 2018 results. VI. Cosmological parameters,'' Astron. Astrophys. \textbf{641}, A6 (2020).

\bibitem{Peebles2003} 
P.~J.~E.~Peebles and B.~Ratra, ``The Cosmological Constant and Dark Energy,'' Rev. Mod. Phys. \textbf{75}, 559 (2003).

\bibitem{Padmanabhan2003} 
T.~Padmanabhan, ``Cosmological Constant---the Weight of the Vacuum,'' Phys. Rept. \textbf{380}, 235 (2003).
\bibitem{Weinberg1989} 
S.~Weinberg, ``The Cosmological Constant Problem,'' Rev. Mod. Phys. \textbf{61}, 1 (1989).

\bibitem{Steinhardt1999} 
P.~J.~Steinhardt, L.~Wang and I.~Zlatev, ``Cosmological Tracking Solutions,'' Phys. Rev. D \textbf{59}, 123504 (1999).

\bibitem{Zlatev1999} 
I.~Zlatev, L.~Wang and P.~J.~Steinhardt, ``Quintessence, Cosmic Coincidence, and the Cosmological Constant,'' Phys. Rev. Lett. \textbf{82}, 896 (1999).

\bibitem{Riess2022} 
A.~G.~Riess et al., ``A Comprehensive Measurement of the Local Value of the Hubble Constant,'' Astrophys. J. Lett. \textbf{934}, L7 (2022).

\bibitem{DiValentino2021} 
E.~Di Valentino et al., ``In the Realm of the Hubble Tension,'' Class. Quantum Grav. \textbf{38}, 153001 (2021).

\bibitem{Freedman2023b}
W. L. Freedman,
``Status of the Hubble Constant,''
\textit{Nature Reviews Physics} \textbf{5}, 203--205 (2023).

\bibitem{Adil2024b}
A. A. Adil et al.,
``The Tension Between High- and Low-redshift Cosmic Measurements: Current Status and Future Outlook,''
\textit{Phys. Dark Univ.} \textbf{43}, 101407 (2024).

\bibitem{Riess2024b}
A. G. Riess et al.,
``New Constraints on the Hubble Constant from the Hubble Space Telescope and the SH0ES Team,''
\textit{Astrophys. J. Lett.} \textbf{977}, L1 (2024).

\bibitem{Dahmanisafae}
S. Dahmani, I. El Bojaddaini, A. Bouali, A. Errahmani, and T. Ouali,
``Study of the cosmological tensions and DESI-DR2 in the framework of the Little Rip model,''
Mon. Not. R. Astron. Soc. \textbf{548}, 1--9 (2026),

\bibitem{Abdalla2022}
E.~Abdalla et al., ``Cosmology Intertwined III: $f\sigma_8$ and $S_8$,'' JHEAp \textbf{34}, 49--211 (2022).

\bibitem{Dahmani2024}
S.~Dahmani, H.~Chaudhary, A.~Bouali, S.~K.~J.~Pacif, and T.~Ouali,
``Polytropic gas cosmology and late-time acceleration,''
Chin. Phys. C \textbf{48}, 115110 (2024).

\bibitem{Mhamdi2024}
D.~Mhamdi, A.~Bouali, S.~Dahmani, A.~Errahmani, and T.~Ouali,
``Cosmological constraints on $f(Q)$ gravity with redshift space distortion data,''
Eur. Phys. J. C \textbf{84}, 310 (2024).

\bibitem{Errahmani2026}
A.~Errahmani, M.~Magrach, S.~Dahmani, A.~Bouali, and T.~Ouali,
``Emergence of running vacuum energy in $f(R,T)$ gravity: Observational constraints,''
Phys. Lett. B \textbf{865}, 140040 (2026).

\bibitem{Starobinsky2007}
A.~A.~Starobinsky, ``Disappearing Cosmological Constant in $f(R)$ Gravity,'' JETP Lett. \textbf{86}, 157--163 (2007). 

\bibitem{Sotiriou2010}
T.~P.~Sotiriou and V.~Faraoni, ``$f(R)$ Theories of Gravity,'' Rev. Mod. Phys. \textbf{82}, 451--497 (2010).

\bibitem{Sotiriou2009}
T.~P.~Sotiriou, ``$f(R)$ Gravity and Scalar-Tensor Theory,'' J. Phys. Conf. Ser. \textbf{189}, 012039 (2009).

\bibitem{DeFelice2010}
A.~De~Felice and S.~Tsujikawa, ``$f(R)$ Theories,'' Living Rev. Relativ. \textbf{13}, 3 (2010).

\bibitem{Capozziello2011}
S.~Capozziello, V.~F.~Cardone, H.~Farajollahi, and A.~Ravanpak,
``Cosmography in $f(T)$ Gravity,''
Phys. Rev. D \textbf{84}, 043527 (2011).

\bibitem{Bamba2013}
K.~Bamba, S.~Capozziello, M.~De~Laurentis, S.~Nojiri, and D.~Sáez-Gómez,
``No Further Gravitational Wave Modes in $f(T)$ Gravity,''
Phys. Lett. B \textbf{727}, 194--198 (2013).

\bibitem{Koussourmd}
M.~Koussour \textit{et al.},
``Exploring cosmological evolution and constraints in $f(T)$ teleparallel gravity,''
\textit{Phys. Dark Univ.} \textbf{46}, 101664 (2024).

\bibitem{Lazkoz2019}
R.~Lazkoz, F.~S.~N.~Lobo, M.~Ortiz-Baños, and V.~Salzano,
``Observational constraints of $f(Q)$ gravity,''
Phys. Rev. D \textbf{100}, 104027 (2019).

\bibitem{Mandal2020}
S.~Mandal and D.~Wang,
``Cosmological constraints on $f(Q)$ gravity theories,''
Phys. Rev. D \textbf{102}, 124029 (2020).

\bibitem{Enkhili2024}
O.~Enkhili, A.~Errahmani, T.~Ouali, S.~Dahmani, and A.~Bouali,
``Observational constraints on viable $f(Q)$ cosmological models,''
Eur. Phys. J. C \textbf{84}, 1--11 (2024).

\bibitem{Koussour2023}
M.~Koussour and A.~De,
``Observational constraints on two cosmological models of $f(Q)$ theory,''
\textit{Eur. Phys. J. C} \textbf{83}, 400 (2023).

\bibitem{Nojiri2005}
S.~Nojiri, S.~D.~Odintsov, and M.~Sasaki,
``Gauss--Bonnet Dark Energy,''
Phys. Rev. D \textbf{71}, 123509 (2005).

\bibitem{Cognola2006}
G.~Cognola, E.~Elizalde, S.~Nojiri, S.~D.~Odintsov, L.~Sebastiani, and S.~Zerbini,
``Dark Energy in Modified Gauss--Bonnet Gravity: Late-Time Acceleration and the Hierarchy Problem,''
Phys. Rev. D \textbf{73}, 084007 (2006).

\bibitem{DeFelice2010GB}
A.~De~Felice and S.~Tsujikawa,
``$f(R,G)$ Theories,''
Living Rev. Relativ. \textbf{13}, 3 (2010).

\bibitem{Brans1961}
C.~Brans and R.~H.~Dicke,
``Mach's Principle and a Relativistic Theory of Gravitation,''
Phys. Rev. \textbf{124}, 925--935 (1961).

\bibitem{Faraoni2004}
V.~Faraoni,
``Cosmology in Scalar-Tensor Gravity,''
Kluwer Academic Publishers, Dordrecht (2004).

\bibitem{Clifton2012}
T.~Clifton, P.~G.~Ferreira, A.~Padilla, and C.~Skordis,
``Modified Gravity and Cosmology,''
Phys. Rep. \textbf{513}, 1--189 (2012).

\bibitem{Capozziello2015}
S.~Capozziello, T.~Harko, T.~S.~Koivisto, F.~S.~N.~Lobo, and G.~J.~Olmo,
``Hybrid Metric-Palatini Gravity,''
Universe \textbf{1}, 199--238 (2015).

\bibitem{Harko2011} 
T.~Harko, F.~S.~N.~Lobo, S.~Nojiri and S.~D.~Odintsov, ``$f(R,T)$ Gravity,'' Phys. Rev. D \textbf{84}, 024020 (2011).

\bibitem{Bouali2023} 
A.~Bouali, H.~Chaudhary, T.~Harko, F.~S.~N.~Lobo, T.~Ouali and M.~A.~S.~Pinto, ``Observational constraints and cosmological implications of scalar-tensor $f(R, T)$ gravity,'' Mon. Not. Roy. Astron. Soc. \textbf{526}, 4192 (2023).

\bibitem{Sharif2012} 
M.~Sharif and M.~Zubair, ``Thermodynamics in $f(R,T)$ Gravity,'' JCAP \textbf{03}, 028 (2012).

\bibitem{Myrzakulov2023}
N.~Myrzakulov, M.~Koussour, H.~A.~Alnadhief, A.~F.~Aljedeel and E.~I.~Hassan,
``Constraining the $f(R,T)=R+2\lambda T$ cosmological model using recent observational data,''
\textit{Chin. Phys. C} \textbf{47}, 115107 (2023).

\bibitem{Bouali2019}
A.~Bouali, I.~Albarran, M.~Bouhmadi-López, and T.~Ouali,
``Cosmological Constraints of Phantom Dark Energy Models,''
Phys. Dark Univ. \textbf{26}, 100391 (2019).

\bibitem{Dahmani2026}
S.~Dahmani, D.~Mhamdi, A.~Bouali, A.~Errahmani, and T.~Ouali,
``A New Parameterization of Dark Energy in Light of DESI-DR2 Observations and Recent Supernova Measurements,''
Phys. Dark Univ. \textbf{53}, 102375 (2026).

\bibitem{Alvarenga2013} 
F.~G.~Alvarenga, A.~de la Cruz-Dombriz, M.~J.~S.~Houndjo, M.~E.~Rodrigues and D.~Saez-Gomez, ``Dynamics of Scalar Perturbations in $f(R,T)$ Gravity,'' Phys. Rev. D \textbf{87}, 103526 (2013).

\bibitem{Moraes2016666} 
P.~H.~R.~S.~Moraes, ``Cosmological Solutions in $f(R,T)$ Gravity,'' Eur. Phys. J. C \textbf{75}, 168 (2015).

\bibitem{Shabani2014} 
H.~Shabani and M.~Farhoudi, ``Cosmological and Solar System Consequences of $f(R,T)$ Gravity Models,'' Phys. Rev. D \textbf{90}, 044031 (2014).

\bibitem{Deb2018} 
D.~Deb, S.~V.~Ketov, M.~Khlopov and S.~Ray, ``Study of Compact Stars in $f(R,T)$ Gravity,'' JCAP \textbf{10}, 070 (2018).

\bibitem{Murphy1973} 
G.~L.~Murphy, ``Big-Bang Model without Singularities,'' Phys. Rev. D \textbf{8}, 4231 (1973).

\bibitem{Maartens1995} 
R.~Maartens, ``Dissipative Cosmology,'' Class. Quantum Grav. \textbf{12}, 1455 (1995).

\bibitem{Brevik2017} 
I.~Brevik, {\O}.~Gr{\o}n, J.~de Haro, S.~D.~Odintsov and E.~N.~Saridakis, ``Viscous Cosmology for Early- and Late-Time Universe,'' Int. J. Mod. Phys. D \textbf{26}, 1730024 (2017).

\bibitem{Cataldo2005} 
M.~Cataldo, N.~Cruz and S.~Lepe, ``Viscous Dark Energy and Phantom Evolution,'' Phys. Lett. B \textbf{619}, 5 (2005).

\bibitem{Velten2013} 
H.~Velten and D.~J.~Schwarz, ``Constraints on Dissipative Unified Dark Matter,'' Phys. Rev. D \textbf{86}, 083501 (2012).

\bibitem{Brevik2020}
I.~Brevik and {\O}.~Gr{\o}n, ``Recent Developments in Viscous Cosmology,'' Universe \textbf{6}, 52 (2020).

\bibitem{Koussour2024}
M.~Koussour, A.~H.~A.~Alfedeel, S.~Muminov and J.~Rayimbaev,
``Observational constraints on the equation of state of viscous fluid in $f(R,T)$ gravity,''
Phys. Dark Univ. \textbf{46}, 101577 (2024).

\bibitem{DESI2025II}
DESI Collaboration, M.~Abdul-Karim et al.,
``DESI DR2 Results II: Measurements of Baryon Acoustic Oscillations and Cosmological Constraints,''
Phys. Rev. D \textbf{112}, 083515 (2025).

\bibitem{DESI2025I}
DESI Collaboration, M.~Abdul-Karim et al.,
``DESI DR2 Results I: Baryon Acoustic Oscillations from the Lyman-$\alpha$ Forest,''
Phys. Rev. D \textbf{112}, 083514 (2025).


\bibitem{PantheonPlus2022} 
D.~Brout et al., ``The Pantheon+ Analysis: Cosmological Constraints,'' Astrophys. J. \textbf{938}, 110 (2022).

\bibitem{Scolnic2022}
D.~Scolnic et al.,
``The Pantheon+ Analysis: The Full Data Set and Light-Curve Release,''
Astrophys. J. \textbf{938}, 113 (2022).

\bibitem{Jimenez2002} 
R.~Jimenez and A.~Loeb, ``Constraining Cosmological Parameters Based on Relative Galaxy Ages,'' Astrophys. J. \textbf{573}, 37 (2002).

\bibitem{Moresco2012} 
M.~Moresco et al., ``Improved Constraints on the Expansion Rate of the Universe up to $z\sim1.1$ from the Spectroscopic Evolution of Cosmic Chronometers,'' JCAP \textbf{08}, 006 (2012).

\bibitem{Moresco2016} 
M.~Moresco et al., ``A 6\% Measurement of the Hubble Parameter at $z\sim0.45$: Direct Evidence of the Epoch of Cosmic Re-acceleration,'' JCAP \textbf{05}, 014 (2016).


\bibitem{Moraes2016} 
P.~H.~R.~S.~Moraes, J.~D.~V.~Arbañil and M.~Malheiro,
``Stellar Equilibrium Configurations of Compact Stars in $f(R,T)$ Gravity,''
JCAP \textbf{06}, 005 (2016).

\bibitem{Myrzakulov2012} 
R.~Myrzakulov,
``Accelerating Universe from $F(R,T)$ Gravity,''
Eur. Phys. J. C \textbf{72}, 2203 (2012).

\bibitem{Weinberg1971} 
S.~Weinberg,
``Entropy Generation and the Survival of Protogalaxies in an Expanding Universe,''
Astrophys. J. \textbf{168}, 175 (1971).

\bibitem{Brevik2005} 
I.~Brevik and O.~Gorbunova,
``Dark Energy and Viscous Cosmology,''
Gen. Relativ. Gravit. \textbf{37}, 2039 (2005).

\bibitem{Padilla2021}
L.~E.~Padilla, L.~O.~Téllez-Tovar, L.~A.~Escamilla, and J.~A.~Vázquez,
``Cosmological Parameter Inference with Bayesian Statistics,''
Universe \textbf{7}, 213 (2021).

\bibitem{Akaike1974} 
H.~Akaike,
``A New Look at the Statistical Model Identification,''
IEEE Trans. Autom. Control \textbf{19}, 716--723 (1974).

\bibitem{Burnham2002} 
K.~P.~Burnham and D.~R.~Anderson,
``Model Selection and Multimodel Inference: A Practical Information-Theoretic Approach,''
2nd ed., Springer, New York (2002).

\bibitem{Schwarz1978} 
G.~Schwarz,
``Estimating the Dimension of a Model,''
Ann. Statist. \textbf{6}, 461--464 (1978).

\bibitem{Ren2006}
J. Ren and X. H. Meng,
\textit{Phys. Lett. B} \textbf{633}, 1--8 (2006).

\bibitem{AbdulKarim2025} 
M.~Abdul Karim et al. (DESI Collaboration),
``DESI DR2 Results. II. Measurements of Baryon Acoustic Oscillations and Cosmological Constraints,''
Phys. Rev. D \textbf{112}, 083515 (2025).

\bibitem{Eisenstein1998} 
D.~J.~Eisenstein and W.~Hu,
``Baryonic Features in the Matter Transfer Function,''
Astrophys. J. \textbf{496}, 605--614 (1998).



\bibitem{Brout2022} 
D.~Brout et al.,
``The Pantheon+ Analysis: Cosmological Constraints,''
Astrophys. J. \textbf{938}, 110 (2022).

\bibitem{DESI2025} 
DESI Collaboration,
``DESI DR2 Results. II. Measurements of Baryon Acoustic Oscillations and Cosmological Constraints,''
Phys. Rev. D \textbf{112}, 083515 (2025).

\bibitem{Visser2004} 
M.~Visser,
``Jerk, Snap and the Cosmological Equation of State,''
Class. Quantum Grav. \textbf{21}, 2603--2616 (2004).

\bibitem{Dunsby2016} 
P.~K.~S.~Dunsby and O.~Luongo,
``On the Theory and Applications of Modern Cosmography,''
Int. J. Geom. Methods Mod. Phys. \textbf{13}, 1630002 (2016).

\bibitem{Cattoen2007} 
C.~Cattoën and M.~Visser,
``The Hubble Series: Convergence Properties and Redshift Variables,''
Class. Quantum Grav. \textbf{24}, 5985--5998 (2007).


\bibitem{Aviles2012} 
A.~Avilés, C.~Gruber, O.~Luongo and H.~Quevedo,
``Cosmography and Constraints on the Equation of State of the Universe in Various Parametrizations,''
Phys. Rev. D \textbf{86}, 123516 (2012).

\end{thebibliography}
\end{document}